\documentstyle[fleqn,epsfig]{aipproc}

\begin{document}
\title{Neutral Hyperon and \\ H-Dibaryon Results from KTeV}

\author{Nickolas Solomey}
\address{Enrico Fermi Institute, The University of Chicago, Illinois
60637}

\maketitle

\begin{abstract}
Results from the KTeV experiment at Fermilab using
$\Xi^0$ hyperons are presented, 
especially the first form-factor (FF) measurement from its semi-leptonic 
decay. This decay, which is an important test of the Standard Model, was 
observed for the first time using the 1997 KTeV experiment.
We also report on the analysis of a large samples of $\Xi^0$ weak
radiative
decays, and a search for the decay of a lightly bound $H^0$ dibaryon.
\end{abstract}

\section*{Introduction}
The KTeV experiment is a neutral beam experiment produced by 
protons from the Tevatron
accelerator at 800 GeV/c impinging on a target at a 4.8 mrad angle. The
fiducial decay volume starts 90 m down stream of a target because of the
space needed to collimate the neutral beam which is where
most of the neutral particles decay and is also the location of
the sweeping magnetics that eliminate the charged
particles. The decay volume
from 90 to 160 m from the target is an ultra high vacuum to reduce 
interactions and has scintillator ring counters to veto those events where
the
particles have left the fiducial volume.
A spectrometer consisting of tracking chambers, an analysis magnet,
electromagnetic calorimetry (CsI) \cite{roodman}, particle identification 
by transition radiation detectors (TRD) \cite{solomey}, and a muon counter 
system with 5 m of iron filter directly follows the decay volume. 

Data was collected in 16 triggers for two different experimental
configurations in 1997 and
1999. A rare kaon decay program E799 \cite{sashaL,eva}, and the search for
direct 
CP violation E832 \cite{sashaG} are presented elsewhere in these
proceedings.
Presented here are part of the results from a neutral hyperon program that
had three triggers in the E799 experiment configuration, 
and limited to the results from the 1997 run.

\section*{Physics Analysis}
Physics results from the KTeV hyperon program are:
\vspace{0.15cm}
\paragraph*{Semi-leptonic Decays:}
Neutral hyperon semi-leptonic decays accessible in KTeV are the
beta-decay $\Xi^0 \rightarrow \Sigma^+ \hspace{0.1cm} e^{-} \hspace{0.1cm}
\bar{\nu}_{e}$, and muon decay $\Xi^0 \rightarrow \Sigma^+ \hspace{0.1cm} 
\mu^- \hspace{0.1cm} \bar{\nu}_{\mu}$.
They are important to study for their weak decay FF which give an 
understanding of their underlying structure. 
In the V-A formulation the transition amplitude for the beta decay is:
\begin{equation}M = \frac{G}{\sqrt{2}}
<\Sigma|J^{\lambda}|\Xi>{\bar{u}_{e}} 
\gamma_{\lambda} (1 + \gamma_{5}) u_{\nu} \end{equation}
The V-A hadronic current can be written as:
\begin{eqnarray*} <\Sigma|J^{\lambda}|\Xi> = {\cal C} \hspace{0.1cm} i 
\hspace{0.2cm}
\bar{u}(\Sigma) & [ & f_{1}\gamma^{\lambda}
+ f_{2} \frac{\sigma^{\lambda \upsilon}\gamma_{\upsilon}}{M_{\Xi}} +
f_{3} q^{\lambda} \frac{M_e}{M_{\Xi}} + \\
& [ & g_{1} \gamma^{\lambda} + g_{2} 
\frac{\sigma^{\lambda \upsilon} \gamma_{\upsilon}}{M_{\Xi}} + 
g_{3} q^{\lambda}  \frac{M_e}{M_{\Xi}} ]
\hspace{0.1cm} \gamma_{5} \hspace{0.25cm} ] u(\Xi) 
\hspace{1.9cm} (2) \end{eqnarray*}
where ${\cal C}$ is the CKM matrix element, and $q$ is the momentum
transfer.
There are 3 vector FF: $f_1$ (vector), $f_2$ (weak magnetism) and
$f_3$ (an induced scaler); plus 3 axial-vector FF: $g_1$ (axial vector),
$g_2$ (weak electricity) and $g_3$ (an induced pseudo-scaler). All six
FF are real if time reversal invariance is valid. The quark
model predicts a nonzero but small $g_2$ FF if SU(3) breaking is
sizable, but the standard model assumes this FF is zero. The
FF $f_3$ and $g_3$ for the beta decay, are expected to be large
(i.e. $\frac{g_3}{g_1}\sim$8), but it is multiplied by
$\frac{M_e}{M_{\Xi}}$
making this term negligably small so as not to contribute any noticeable 
effect, but for the muon decay this may no longer be assumed. Furthermore,
neither of these decays had previously been observed so measuring their
branching ratio was also important as a test of the standard model, and
in the case of the muon decay this could be the first place to look
for a $g_3$ FF. The final results for the beta decay are a 
branching ratio of (2.60~$\pm$0.11~$\pm$0.16)~$\times$~10$^{-4}$,
based on 626 events where the first error is statistical and the second 
systematic, and the theoretical expected is 2.6~$\times$~10$^{-4}$.
For the muon decay its preliminary branching ratio is
(3.5~$^{+2.0}_{-1.0}$~$^{+0.5}_{-1.0}$)~$\times$~10$^{-6}$ based on 5
events,
again the first error is statistical and the second systematic, while
the asymmetric error bars are from the small number of events and poisson
statistics at the 68\% C.I. \cite{feldman}; theoretical expected is
2.6~$\times$~10$^{-6}$.
\begin{figure}[h]
\begin{center}
\mbox{
\epsfig{file=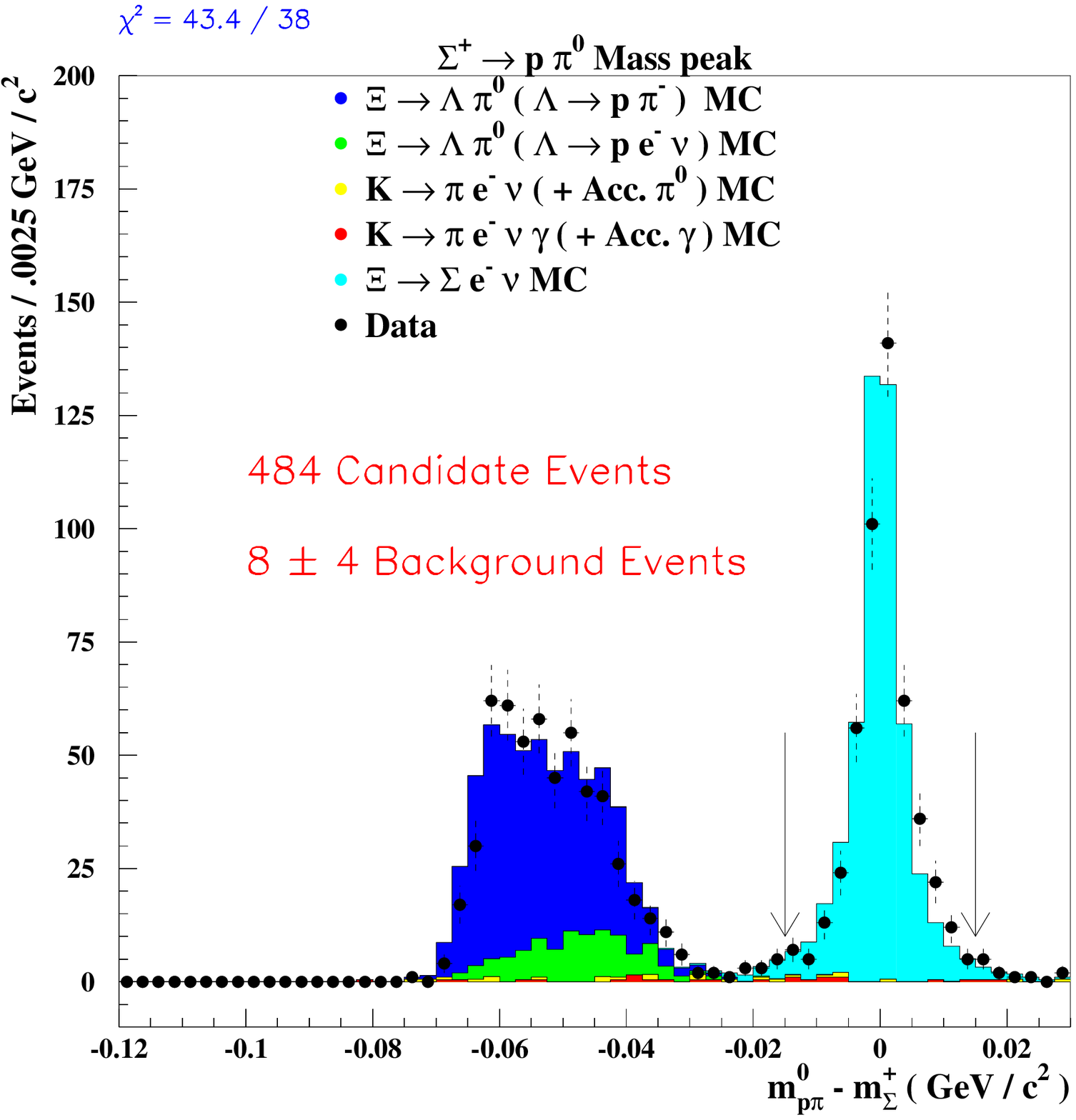,%
height=0.47\linewidth}
}
\mbox{
\epsfig{file=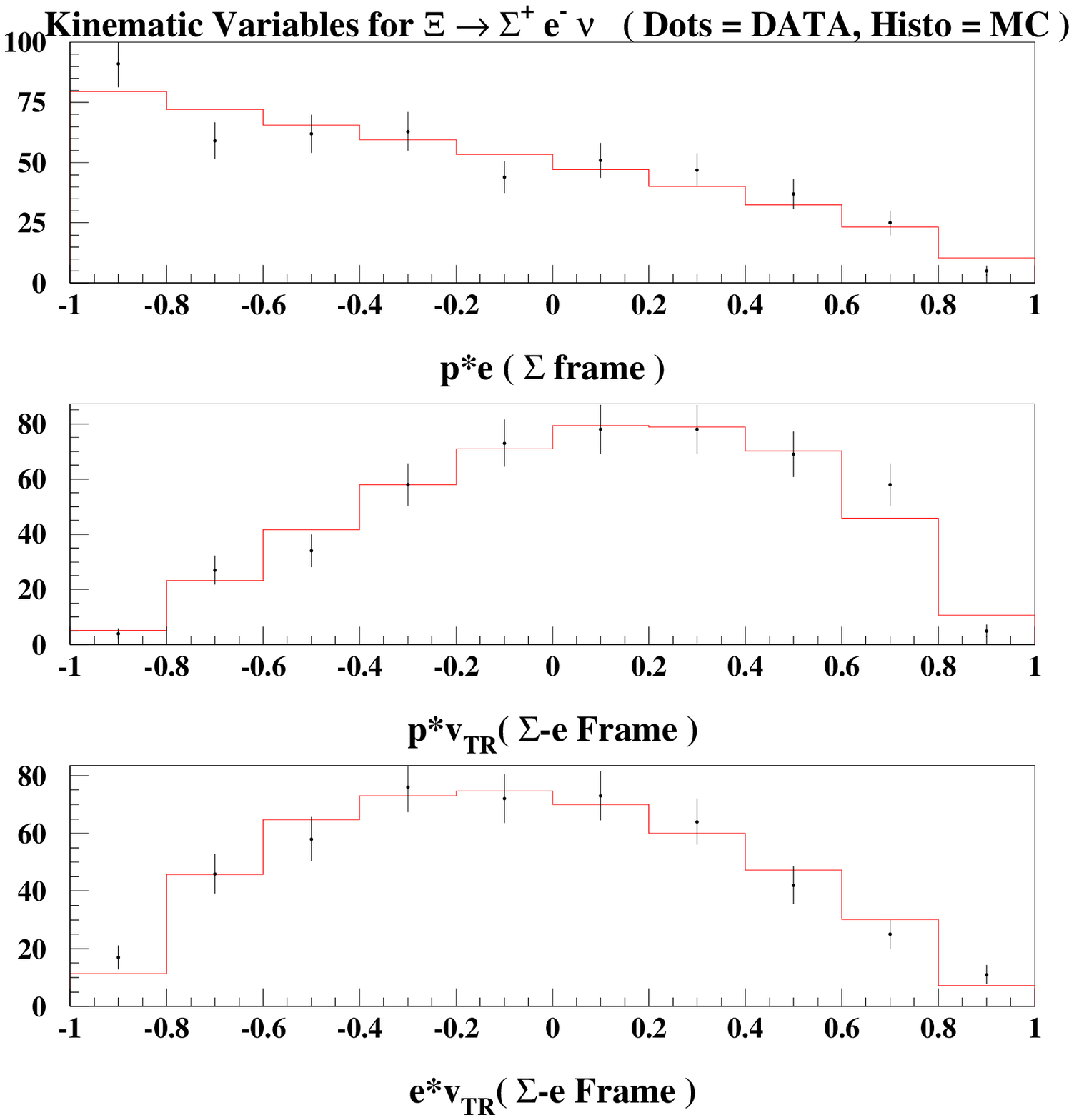,%
height=0.47\linewidth}
}
\end{center}
\caption{On the left is the $\Xi^0$ beta decay compared to its 
normalized MC and possible backgrounds, and on the right is the
angular distribution in the $\Sigma^+$ reference frame of the
angular asymmetry of the proton to electron, reconstructed 
neutrino and between the reconstructed neutrino and the electron.}
\label{beta}
\end{figure}

\begin{figure}
\begin{center}
\mbox{
\epsfig{file=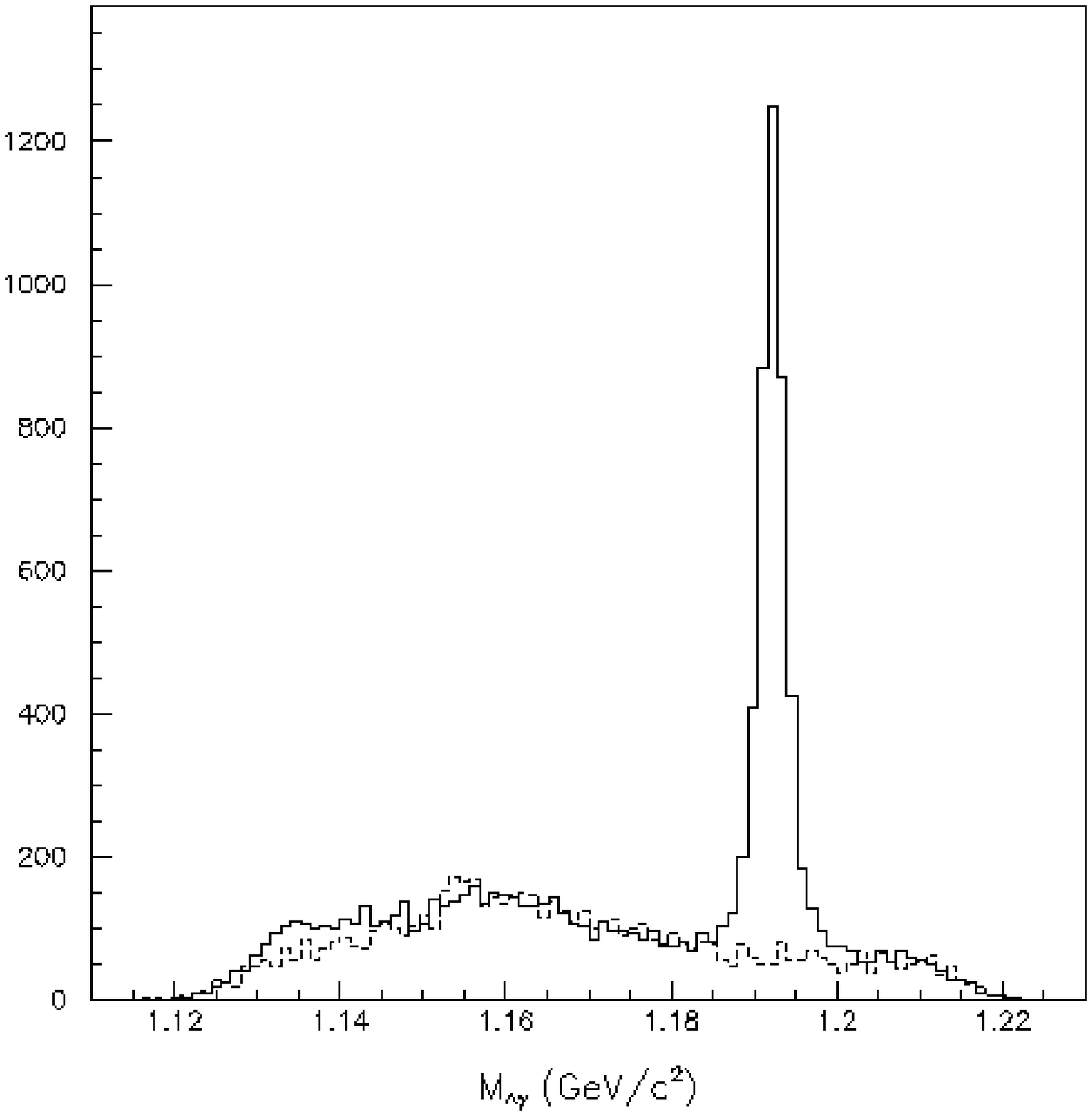,%
height=0.31\linewidth}
}
\mbox{
\epsfig{file=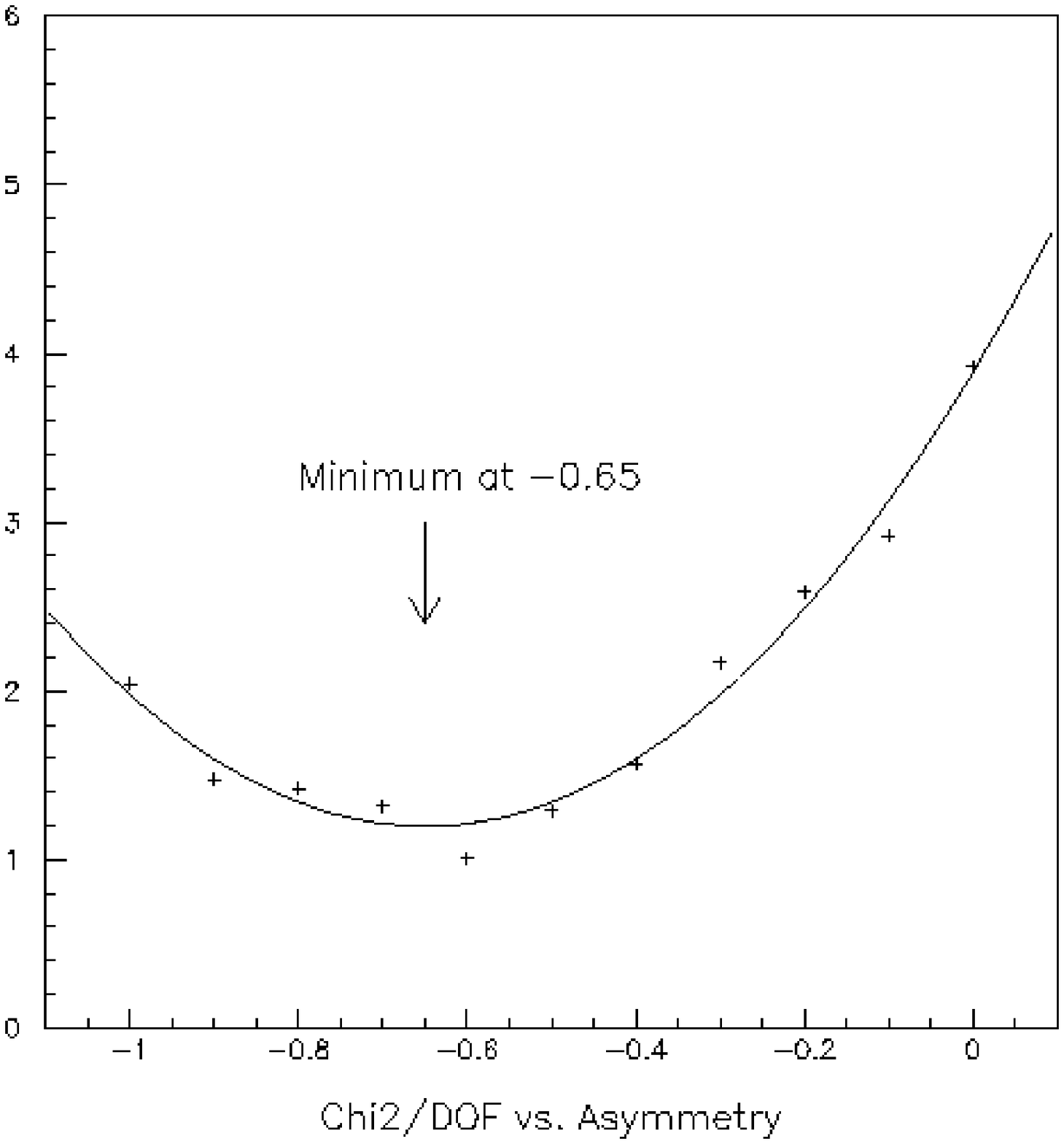,%
height=0.31\linewidth}
}
\end{center}
\caption{On the left is the $\Xi^0 \rightarrow \Sigma^0 \gamma$ decay
signal 
compared to its normalized MC, and on the
right the determination of the gamma angular asymmetry.}
\label{wrhd}
\end{figure}

A very clean sample of $\Xi^0$ beta decays, see figure \ref{beta} left,
was obtained by using the TRD detector. The decays of the $\Sigma^+$
has a 98\% analyzing power, and this fact makes its equivalent to a fully
polarized beam. However, spin alignment magnetics gave
the ability to control this and then test the technique on the much 
bigger normal mode decays: $\Xi^0 \rightarrow \Lambda^0 \pi^0$. By working
in the $\Sigma^+$ reference frame all of the FF could be
devised by measuring the angular distribution of the proton relative
to the electron, neutrino (we typically use the reconstructed 
transverse neutrino direction)
see figure \ref{beta} right, as well as test the technique by comparing
the
proton direction relative to the reconstructed $\Xi^0$. The final four 
FF are: $f_1 = 0.99 \pm 0.14$, $\frac{g_1}{f_1}=1.24 \pm 0.27$,
$\frac{f_2}{f_1}=2.3 \pm 1.3$, and $\frac{g_2}{f_1}= -1.4 \pm 2.1$;
here this analysis used the previously quoted branching ratio, and
permitted
the $g_2$ FF to float. The $g_2$ FF is consistent with
zero and in another analysis it was constrained to be zero and the
FF reanalyzed and they essentially remained unchanged.
\vspace{0.15cm}
\paragraph*{Radiative Decays:}
Neutral hyperon radiative decays accessible in KTeV are:
$\Xi^0 \rightarrow \Sigma^0 \gamma$ and
$\Xi^0 \rightarrow \Lambda^0 \hspace{0.1cm} \gamma$. 
Although these decays are not simple first-order processes but 
actually proceed
threw complicated diagrams, they are important to study because
they are difficult to calculate, but yet have high
branching ratios. Furthermore, the angular emission of the high
energy gamma relative to the hyperon polarization can be related
to the underlying mechanism of its decay as a test of various 
theories. The decay $\Xi^0 \rightarrow \Sigma^+ \gamma$ in KTeV is 
identified when a second
radiative decay of $\Sigma^0 \rightarrow \Lambda^0 \gamma$ and 
then for both radiative decays of the $\Xi^0$ 
through the charged particle decay of $\Lambda^0 \rightarrow p^+ \pi^-$. 
As with the semi-leptonic decays the self analyzing power with hyperon 
decays permits the polarization to be determined relative to the gamma
angular 
distribution. The final results for the 
radiative decay $\Xi^0 \rightarrow \Sigma^0 \gamma$ is a 
branching ratio of (3.34~$\pm$0.05~$\pm$0.11)~$\times$~10$^{-3}$,
based on 4048 events, shown in figure \ref{wrhd} left, 
where the first error is statistical and the second 
systematic. The angular
distribution of the gamma relative to the $\Xi^0$ polarization vector;
is -0.65 $\pm$0.13, see figure \ref{wrhd} right.
The radiative decay $\Xi^0 \rightarrow \Lambda^0 \gamma$ 
preliminary branching ratio is 
(0.95~$\pm$0.03$\pm$0.09)~$\times$~10$^{-3}$ based on 1105 events.
\vspace{0.15cm}
\paragraph*{Dibaryon Search:}
The KTeV hyperon data was used to search for a six quark neutral hadron
state
which has two strange, two up and two down quarks that 
decays $H^0 \rightarrow \Lambda^0 p^+ \pi^-$. The data used for 
normalizing the decay was $\Xi^0 \rightarrow \Lambda^0 \pi^0$ with the
decay of the $\Lambda^0 \rightarrow p^+ \pi^-$ and $\pi^0 \rightarrow e^+ 
e^- \gamma$. These two decays share similar technical details for their
analysis: two different charged particle vertices, a reconstructible
$\Lambda^0$ mass peak and a stiff proton. Using the 1997 data from
KTeV E799 run we have
ruled out the existence of a $H^0$ dibaryon in the mass range from
2.194 to 2.231 GeV/c$^2$ for long lifetimes of 5~$\times$~10$^{-10}$ to
1~$\times$~10$^{-3}$ seconds at the 90\% CI \cite{ram}.

\section*{Conclusions}
The results quoted here are the final measurements for these decays
from the 1997 KTeV run. Three times more data exists from
the 1999 run which is in the process of being analyzed. Future results
are planned to make improvements in all decay modes, and some decays
previously excluded by trigger requirements, such as: $\Xi^0 \rightarrow
\Lambda^0 \pi^0$ where the decay $\pi^0 \rightarrow e^+ e^- \gamma$ will
provide a sample of decays where the z vertex can be measured from
the two electrons for use in a $\Xi^0$ mass and lifetime improved 
measurement, the decay $\Xi^0 \rightarrow \Sigma^0 \gamma$ with the
special decay $\Sigma^0 \rightarrow \Lambda^0 e^+ e^-$, or the 
radiative decay $\Xi^0 \rightarrow \Lambda^0 \pi^0 \gamma$. All of
these decays are especially well suited for the excellent abilities of
$\pi^0$ and $\gamma$ measurements with our high precession CsI 
electromagnetic calorimeter, and our unmatched $\pi^{\pm}$/e$^{\pm}$
rejection obtained with the TRD system.

\end{document}